\documentstyle[aps,preprint,float,prc]{revtex}
\tighten
\input{boxedeps.tex}
\SetRokickiEPSFSpecial
\HideDisplacementBoxes

\newcommand{\G}{\gamma}

\newcommand{\EPEM}{e^+e^-}

\newcommand{\BE}{\begin{equation}}
\newcommand{\EE}{\end{equation}}

\begin{document}
\draft

\title{Bremsstrahlung from Electrons and Positrons in Peripheral
Relativistic Heavy Ion Collisions}
\author{Kai Hencken and Dirk Trautmann}
\address{
Institut f\"ur theoretische Physik der Universit\"at Basel,
Klingelbergstrasse 82, 4056 Basel, Switzerland
}
\author{Gerhard Baur}
\address{
Institut f\"ur Kernphysik (Theorie), Forschungszentrum J\"ulich, 
52425 J\"ulich, Germany
}

\date{March 8, 1999}

\maketitle

\begin{abstract}
We study the spectrum of the bremsstrahlung photons coming from the
electrons and positrons, which are produced in the strong
electromagnetic fields present in peripheral relativistic heavy ion
collisions. We compare different approaches, making use of the exact
pair production cross section in heavy ion collisions as well as the
double equivalent photon approximation.
\end{abstract}


\section{Introduction}
\label{sec_intro}

The cross-section for electron-positron pair production in relativistic
heavy ion collisions is huge, see, e.g., \cite{BaurHT98} for a general
reference, due to the coherence of all the protons in each nucleus. The
$\EPEM$-pairs are produced in an interaction with two quasi-real
photons. On the other hand, the cross-section for bremsstrahlung in
peripheral relativistic heavy ion collisions was found to be small, both
for real \cite{BertulaniB88} and virtual \cite{MeierHTB98}
bremsstrahlung photons.  This is essentially due to the large mass of
the heavy ions. Since the cross section for $\EPEM$ pair production is
so large, of the order of 100 kbarn for RHIC and LHC, one can expect to
see sizeable effects from the bremsstrahlung radiation of these light
mass particles. It is the purpose of this paper to calculate the
cross-section for bremsstrahlung due to this effect and show its main
characteristics. The general theoretical framework is presented in
Sec.~\ref{sec_theory}. Numerical results along with a discussion are
given for RHIC and LHC conditions in Sec.~\ref{sec_results},
Sec.~\ref{sec_conclusion} contains our conclusions.

\section{Theoretical Framework}
\label{sec_theory}

\begin{figure}[tbhp]
\begin{center}
\ForceHeight{4cm}
\BoxedEPSF{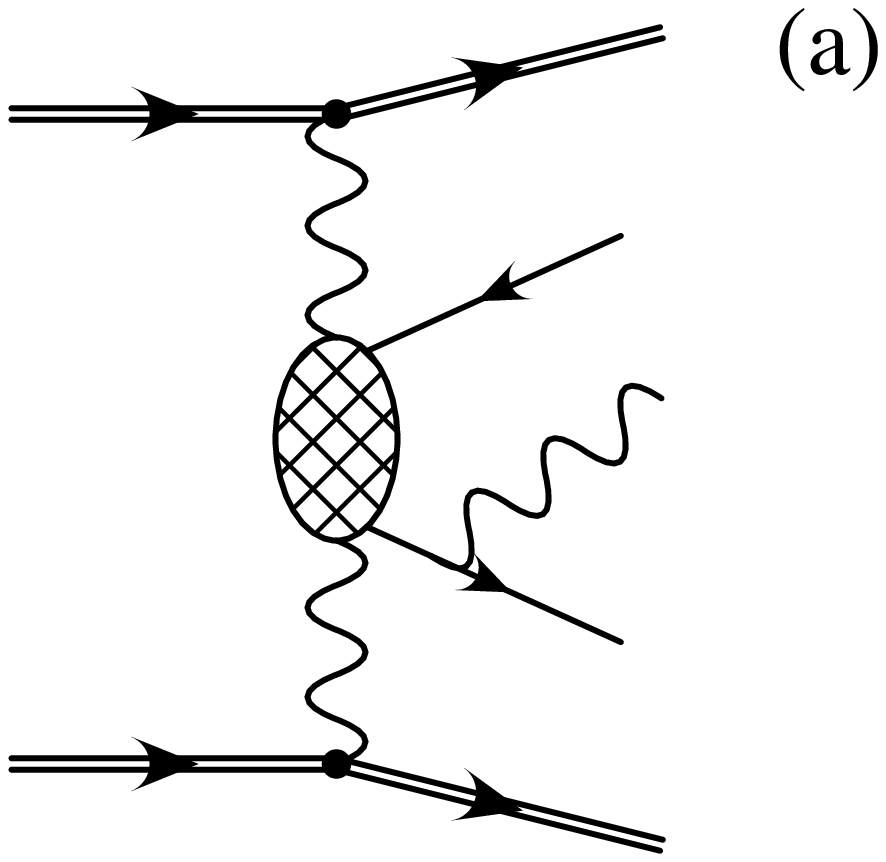}
\ForceHeight{4cm}
\BoxedEPSF{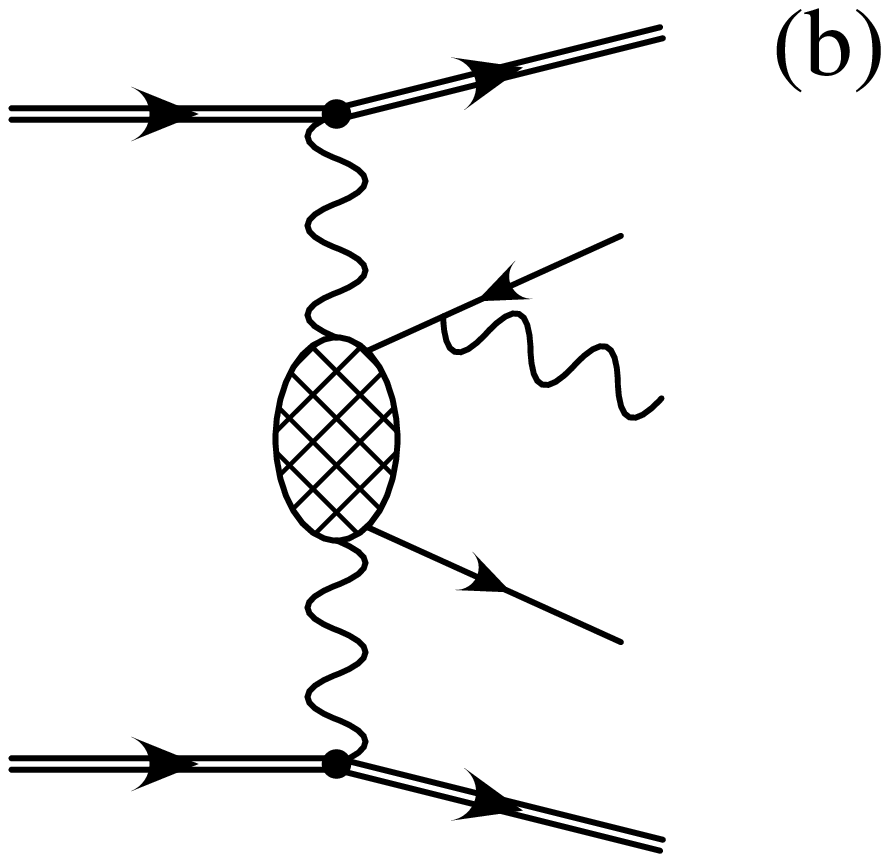}
\ForceHeight{4cm}
\BoxedEPSF{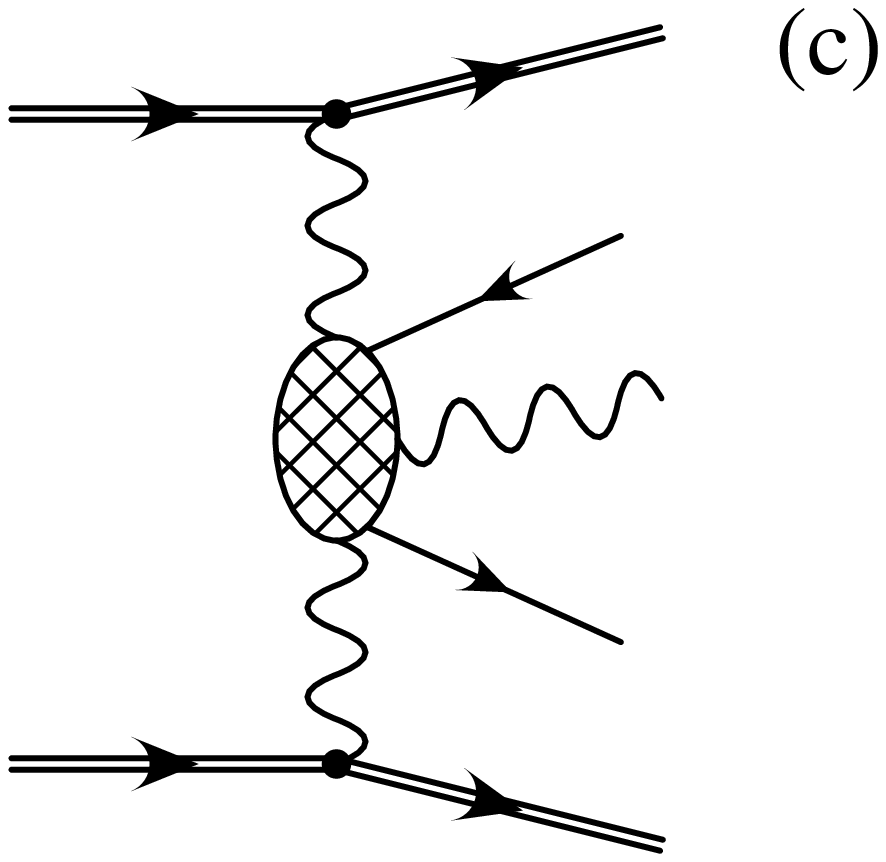}
\end{center}
\caption{\it
The Feynman graphs considered in this paper for the emission of
bremsstrahlung photons in peripheral heavy ion collisions. In the IR
limit, graphs (a) and (b) are dominant.}
\label{fig_graph}
\end{figure}

We consider peripheral collisions where the ions do not interact
strongly with each other (so called ``peripheral collisions''). The
basic mechanism to produce $\EPEM$-pairs is the two-photon mechanism.
In principle the soft bremsstrahlung photons can be emitted from the two
heavy ions, as well as from the produced lepton pair. Due to their large
mass, we can neglect the emission from the heavy ions (see, for example,
\cite{BertulaniB88} and \cite{MeierHTB98}) and we calculate the
bremsstrahlung due to the dileptons only. This is shown in
Fig.~\ref{fig_graph}.  In the low energy limit (infrared (IR) limit) the
photon emission from the external lines is dominant over the emission
from internal lines. This is a well known general result, see, e. g.,
\cite{Weinberg97}, and the cross section for soft photon emission of the
process
\BE
Z + Z \rightarrow Z + Z + e^+ + e^- + \gamma
\EE
can be calculated as
\BE
\frac{d\sigma}{d^3p_+ d^3p_- d\Omega d\omega} =
- e^2 \left[ \frac{p_-}{p_- k} - \frac{p_+}{p_+ k} \right]^2
\frac{\omega^2}{4 \pi^2 \omega} \frac{d\sigma_0}{d^3p_+ d^3p_-},
\label{eq_lowe}
\EE
where $\sigma_0$ denotes the cross section for the $\EPEM$ pair
production in heavy ion collisions. Using the code for the exact lowest
order cross section for the $\EPEM$ pair production in external field
approximation of \cite{AlscherHT97}, we can calculate the soft photon
emission from the outgoing lepton lines numerically according to this
equation. Inclusive bremsstrahlung cross sections are then calculated by
integrating over the unobserved electron pairs
\BE
\frac{d\sigma}{d\Omega d\omega} = \int 
\frac{d\sigma}{d^3p_+ d^3p_- d\Omega d\omega} d^3p_+ d^3p_-.
\label{eq_incl}
\EE
Eq.~(\ref{eq_lowe}) is only valid as long as the energy and momentum of
the bremsstrahlung photon is small compared to the other energies and
momenta involved in the process. Following the discussion in
\cite{LandauLQED}, we therefore use the following restriction condition:

As a measure of the influence of the additionally emitted photon, we use
the invariant mass of the lepton pair with and without the photon
emission. This has the advantage of being a Lorentz scalar. We expect
the low energy approximation to be valid if
\BE
(p_+ + p_- + k)^2 - (p_+ + p_-)^2 \ll (p_+ + p_-)^2,
\label{eq_cond}
\EE
which can be rewritten conveniently as
\BE
2 (p_+ +p_-) \cdot k \ll M_{e+e-}^2,
\label{eq_cond1}
\EE
where $M_{e+e-}$ denotes the invariant mass of the lepton pair.  For a
given value of the photon momentum $k$ this is a restriction on the
values of the momenta $p_+$ and $p_-$ of the lepton pair. In the
numerical calculation, we keep only those processes, where the above
estimate is fulfilled as an inequality.

The majority of the leptons are produced with an invariant mass of the
order of 2--5~$m_e$, see, e.g., \cite{AlscherHT97,BaurF90}.  Therefore
this restriction is not important for photon energies which are much
smaller than 1~MeV.  The momentum of the pair $(p_+ + p_-)$ is also
almost aligned along the beam axis in most cases. As a function of the
angle with the beam axis, the scalar-product $(p_+ + p_-)\cdot k$ will
therefore be smaller for small angles and larger for large angles.
Therefore we expect that this condition is more important for photons
emitted under a large angle from the beam axis.  In the following, we
call this ``{\em method 1}'' to calculate bremsstrahlung cross section.

An alternative approach, to be called ``{\em method 2\/}'', is possible
by using the DEPA (double equivalent photon approximation). The two
heavy ions are accompanied by a spectrum of equivalent (quasireal)
photons.  The collision of these photons leads to
\BE
\gamma + \gamma \rightarrow e^+ + e^- + \gamma.
\label{eq_gg}
\EE
We calculate the cross-section for this subprocess in lowest order QED
using a computer algebra program \cite{form}. This cross-section is then
folded with the equivalent photon numbers $n(\omega)$, which describe
the probability of an equivalent photon to be emitted by one of the ions
leading to
\BE
\frac{d\sigma}{d^3p_+ d^3p_- d\Omega d\omega} =
\int \frac{d\omega_1}{\omega_1}
\int \frac{d\omega_2}{\omega_2} n(\omega_1) n(\omega_2) 
\frac{d\sigma(\gamma+\gamma \rightarrow e^+ e^- + \gamma )}
{d^3p_+ d^3p_-d\Omega d\omega }
\label{eq_fepa}
\EE
This method has the advantage that the kinematics of the bremsstrahlung
emission is treated correctly. There is no restriction on the photon
energy, as is the case for the IR method (eqs
\ref{eq_lowe}--\ref{eq_cond1}).  Of course this calculation can only be
as good as the underlying DEPA used to calculate the $\EPEM$ pair
production.

In addition, even if the DEPA reproduces the total pair production
(only) cross section reasonably well, this cannot be expected
automatically for the bremsstrahlungs cross section, as both processes
are expected to be sensitive to different areas of the $e^+ e^-$ phase
space.  For the equivalent photon spectrum we use (see, e.g.,
\cite{BertulaniB88,JacksonED})
\BE
n(\omega) = \frac{2 Z^2 \alpha}{\pi} [ \xi K_0(\xi) K_1(\xi) - 
\frac{1}{2} \left[ K_1^2(\xi) - K_0^2(\xi) ] \right],
\label{eq_nepa}
\EE 
where the $K_i$ are the McDonald functions, $\xi=\omega R/\gamma$, and
$R$ is a suitably chosen cutoff parameter.  For rather soft pairs ---
which are of importance here --- one may choose $R$ to be of the order
of the Compton wavelength of the electron $\lambda_c = 1/m_e$
\cite{BertulaniB88,JacksonED}.

As {\em ``method 3''} we have also used the low energy (IR)
approximation for the cross-section $d\sigma(\gamma+\gamma \rightarrow
e^+ e^-\gamma)$ in eq.~\ref{eq_fepa}. We use
eqs.~(\ref{eq_lowe})--(\ref{eq_cond1}), where $d\sigma_0$ in
eq.~(\ref{eq_lowe}) is the cross section for the process $\G + \G
\rightarrow e^+ + e^-$.

\begin{figure}
\begin{center}
\ForceHeight{8cm}
\BoxedEPSF{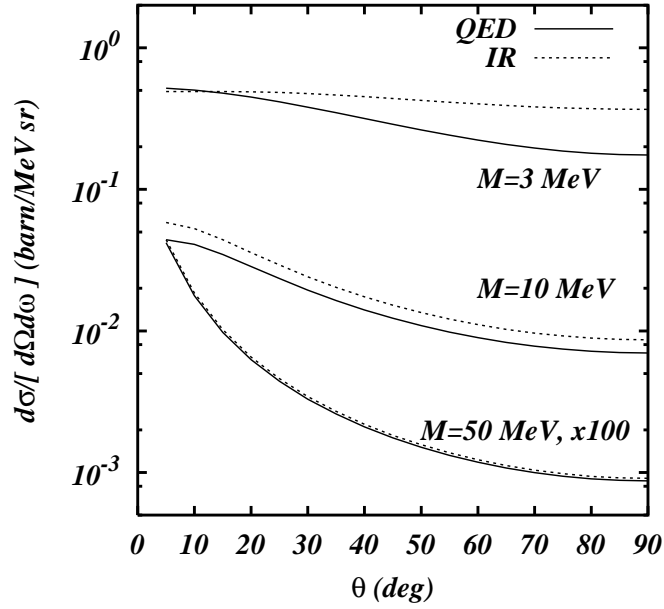}
\end{center}
\caption{The inclusive differential cross section $d\sigma/d\Omega
d\omega$ (see eq.3) for the process(eq.6) is shown for different fixed
photon-photon invariant masses and in the photon-photon rest
frame. $\theta$ is the angle between the bremsstrahlungs photon and one
of the initial photons. The energy of the bremsstrahlung photon is
$\omega$=1 MeV. The solid line are the lowest order QED results, whereas
the dotted line those of the IR approximation.}
\label{fig_ctheta}
\end{figure}

In order to see what range of applicability the IR approximation has, we
have compared the cross section for eq.~(\ref{eq_gg}) using the exact
calculation to the IR method (eqs.~(\ref{eq_lowe})--(\ref{eq_cond1})).
We go to the photon-photon center of mass frame and perform calculations
for different values of the invariant mass. We integrate over the
momenta of the two unobserved leptons.  In this case $d\sigma_0$ denotes
the cross-section for the process $\G + \G \rightarrow e^+ +
e^-$. Figure~\ref{fig_ctheta} shows the cross section $d\sigma/d\Omega
d\omega $, where $\theta$ is the angle between the bremsstrahlung photon
and one of the initial photons.  The energy of the bremsstrahlung photon
was chosen to be $\omega$=1 MeV.  One clearly sees that the IR
approximation is very good for large invariant masses compared to the
bremsstrahlung photon energy. At invariant masses comparable to the
bremsstrahlung photon energy, on the other hand, we find disagreement,
as expected. We also see that the deviation between the exact QED
calculation and the IR approximation is stronger for larger angle, which
is the kind of behavior we expect from the discussion of the restriction
condition (see eqs.~(\ref{eq_cond}) and~(\ref{eq_cond1})).

\section{Numerical Results}
\label{sec_results}

\begin{figure}
\begin{center}
\ForceHeight{8cm}
\BoxedEPSF{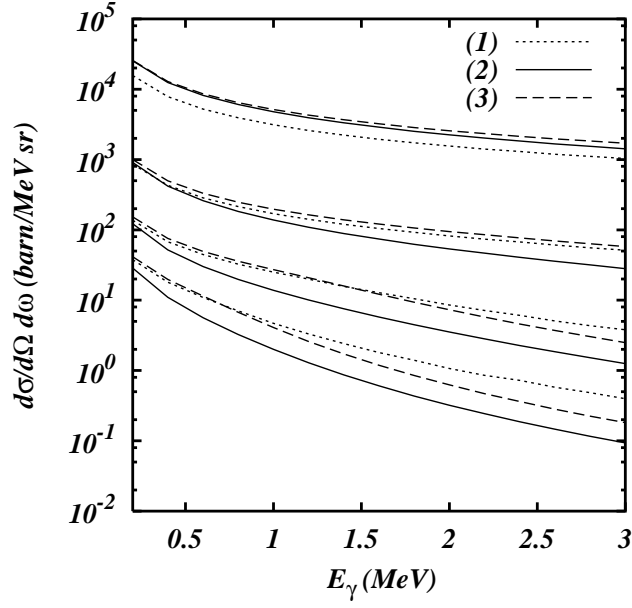}
\end{center}
\caption{The inclusive differential cross section $d\sigma/d\Omega d\omega$ is
shown for RHIC conditions ($\gamma$=100, Au-Au, Z=79). Results are shown for
photon energies up to 3 MeV and for different angles: 1, 10, 30, 90$^0$
from top to bottom. The cutoff parameter $R$ was set to 100fm. The different
calculations ((1)--(3)) are as explained in the text.}
\label{fig_rhic}
\end{figure}

\begin{figure}
\begin{center}
\ForceHeight{8cm}
\BoxedEPSF{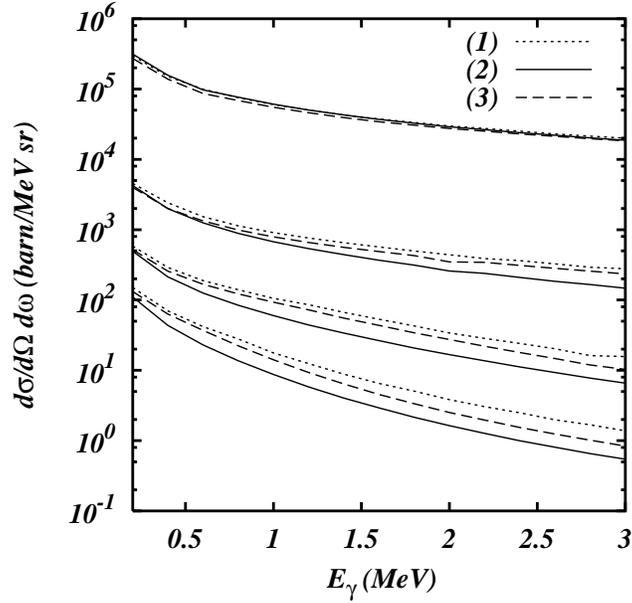}
\end{center}
\caption{Same as figure above, but now for the LHC ($\gamma$=3400,
Pb-Pb, Z=82).}
\label{fig_lhc}
\end{figure}

In the following we use the three different methods as described above
in order to calculate the bremsstrahlung spectrum for the heavy ion
case.  Comparing the different calculations allows us to check both, the
validity of the low energy approximation together with our restriction
condition by comparing method 2 and 3, but also the validity of the DEPA
(by comparing method 1 and 3), which depends on the choice of the cutoff
parameter $R$.  The comparison of the three different results are shown
for the condition at RHIC (Au-Au collisions, $\G=100$) and LHC (Pb-Pb
collisions, $\G=3400$) in Figs.~\ref{fig_rhic} and~\ref{fig_lhc}.  The
two calculations using the DEPA (method 2 and 3) are in agreement with
each other within a factor of two, justifying the use of the IR
approximation (together with the restriction condition
(eq.~(\ref{eq_cond1})).  (Without this condition they would differ by
more than a factor of 100 at 3 MeV and $90^0$.)

Comparing them with the calculation using the exact pair production
(method 1), one finds that using $R$ equal to the Compton wavelength of
the electron, as is usual done, gives cross section, which are too
small. Therefore we have adjusted its value to get better agreement
between them. We get ``fair'' agreement when using $R \approx
100$fm. The results in Figs.~\ref{fig_rhic} and \ref{fig_lhc} are
therefore calculated with this cutoff.  A single $R$ does not seem to be
able to reproduce the results for all angles at the same time. This
situation is similar to the one of pair production with large transverse
momenta (see \cite{stagnoliHBT99}). An improvement, which introduces a
$R$ that depends on the final state, is ``work in progress'' and will be
addressed in a future publication.

We have compared our results also with the analytic expressions as given
in \cite{FadinK73}.  In this reference, photon emission at large angles
was studied for $\EPEM$ collisions. Their approach can be carried over
to the heavy ion case with the appropriate modifications. For relatively
low (equivalent) photon energies, the equivalent photon spectrum of
eq.~(\ref{eq_nepa}) is given approximately by
\BE
n(\omega) \approx \frac{2 Z^2 \alpha}{\pi} \ln\left(
\frac{\gamma}{\omega R}\right). 
\EE
For $\ln(\gamma) \gg \ln(\omega R )$, one obtains
\BE
n(\omega) \approx \frac{2 Z^2 \alpha}{\pi} \ln\left(\gamma\right)
\EE
This can be compared to the expression for the equivalent photon
spectrum used in ref \cite{FadinK73}. In the reference, the expression
\BE
n(\omega)=\frac{Z^2 \alpha}{\pi} L
\label{eq_fadin}
\EE
is used, with $L=2 \ln(2 \gamma) = \ln(s/m_e^2)$ (see their eq.~(3)).
We find the modification of the equivalent photon spectrum due to the
factor $\omega R$ to be important in the heavy ion case (see below). Due
to the low mass of the electrons $\gamma$-values at electron colliders
are much higher than at heavy ion colliders (e.g., at RHIC we have
$\gamma \approx 100$, whereas $\gamma\approx 200 000$ at LEP ($E_{el}
=100$GeV)).

For $K_\perp= \omega \sin(\theta) \gg m_e$ an analytic formula is given
(eq.~(5) of \cite{FadinK73}):
\BE
\frac{d\sigma}{d\Omega d\omega} = \frac{ 4 Z^4 \alpha^5}{\pi^3 K_\perp^4} 
L^2 \left[ \frac{7}{12} L_0 - C \right] \omega
\label{eq_fadin1}
\EE
with $C\approx 0.03$ and $L_0=\ln(K_\perp^2/m_e^2)$.
For $\omega \ll m_e$ one finds (eq. (4) of \cite{FadinK73})
\BE
\frac{d\sigma}{d\Omega d\omega} = \frac{Z^4 \alpha^5}{2 \pi^3 m_e^2} 
\frac{L^2}{\omega \sin^2(\theta)} \left[ \frac{128 \pi^2}{108} -6 \right]
\label{eq_fadin2}
\EE
This equation contains the famous $1/\omega$-dependence of the
bremsstrahlung cross-section. The spectrum therefore extends down to
very low energies, well into the region of visible light. Therefore it
is worthwhile to note, that these photons might also be of interest to
measure/``view'' the interaction region at heavy ion colliders.

\begin{figure}
\begin{center}
\ForceHeight{8cm}
\BoxedEPSF{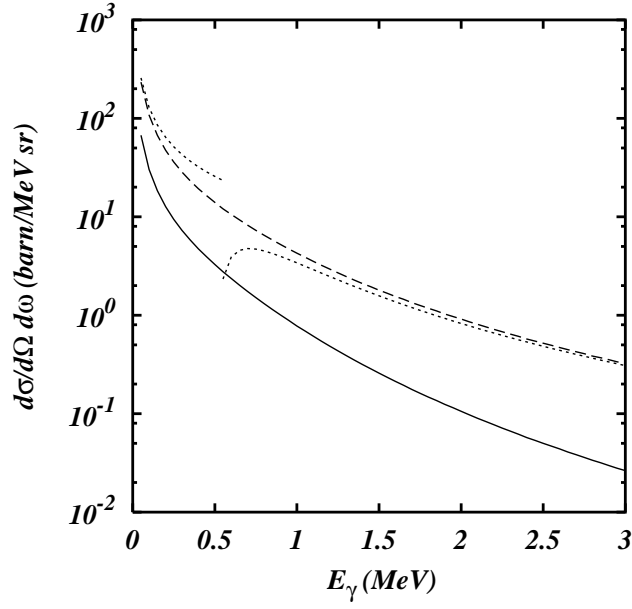}
\end{center}
\caption{The differential cross section $d\sigma/d\Omega d\omega$ is
shown for RHIC (gamma=100,Au-Au,Z=79) conditions and for
$\theta=90^0$. The calculations are done according to method 2. The
solid line is the calculation with the equivalent photon spectrum as
given in eq.~(\protect\ref{eq_nepa}).  The dashed line uses the
simplified equivalent photon spectrum eq.~(\protect\ref{eq_fadin}). The
dotted lines are the analytic expressions of eq.~(4) and~(5) of
\protect\cite{FadinK73} for low and high energies respectively (see eqs
\protect\ref{eq_fadin1} and \protect\ref{eq_fadin2}).}
\label{fig_fadin}
\end{figure}

In Fig.~\ref{fig_fadin} we compare our calculations for $\theta=90^0$
with the one using their equivalent photon spectrum and also with their
analytic expressions eqs.~(\ref{eq_fadin1}) and (\ref{eq_fadin2}).  One
sees a definite discrepancy between the calculations with the different
equivalent photon spectra. This is due to the $\gamma$ values at the
heavy ion colliders, which are not high enough to justify the neglect of
the $\omega R$ term in the equivalent photon spectra (see the discussion
following eq.~(\ref{eq_fadin}) above). One can see also that our
calculation with the simplified version eq.~(\ref{eq_fadin}) of the
equivalent photon spectrum is in very good agreement with the analytical
results for high and low energies of \cite{FadinK73}.

We also note that the analytical formulae (eqs. (\ref{eq_fadin1}) and
(\ref{eq_fadin2})) show a different angular dependence for the high and
low energy limits. For the low energies (eq. (\ref{eq_fadin1})) there is
a $1/(\sin^2\theta)$ dependence, for the high energies (eq
(\ref{eq_fadin1})) a steeper $1/\sin^4\theta$ dependence. This trend is
also visible in our calculations. For larger angles a $1/(\sin^2\theta)$
dependence is seen for bremsstrahlungs photon energies below about
1~MeV, a $1/(\sin^4\theta)$ dependence for larger energies.

It seems interesting to compare these formulae to the ones for
bremsstrahlung due to the heavy ions (see Ch 5.1 of
\cite{BertulaniB88}). These formulae contain a factor
$Z^6\alpha^3/M_A^2$. In the present case we have a scaling factor
$Z^4\alpha^5 /m_e^2$ (or, for higher energies $Z^4\alpha^5
/\omega^2$). As was already discussed qualitatively in the introduction,
the radiation from the electron is more important than the one from the
heavy ions; this can be seen directly from these scaling factors.

\section{Conclusion}
\label{sec_conclusion}

In this paper, we have studied a new type of bremsstrahlung process in
peripheral relativistic heavy ion collisions. We have shown that this is
the dominant mechanism for bremsstrahlung. We have used three different
methods to calculate the corresponding bremsstrahlung spectra. There is
general agreement between these methods in the IR region. We expect the
method 1 to be the most reliable for the soft photon region. For hard
spectra, method 2 is applicable, as it does not rely on the soft photon
approximation; but it depends on the validity of the double equivalent
photon approximation (DEPA).  The DEPA was found to be dependent on the
cutoff parameter $R$, which is theoretically not too well defined. We
find fair agreement between the different approaches by using an $R$,
which is somewhat smaller than conventional wisdom predicts. Method 2
will be most useful for future calculations of the harder part of the
bremsstrahlung spectrum.

These low energy photons constitute a background for relativistic heavy
ion colliders.  Unlike the copiously produced low energy electrons and
positrons, they are of course not bent away by the magnets and could be
a hazard for the detectors.  Recently the soft bremsstrahlung photons
from central ultrarelativistic nucleus-nucleus collisions were suggested
to be used to infer the rapidity distribution of the outgoing charge
\cite{JeonKCS98}. The presently considered soft photons from peripheral
collisions could be a source of background for the considered
experiment.
%
%
%


\begin{references}

\bibitem{BaurHT98}
G. Baur, K. Hencken, and D. Trautmann, Topical Review, 
J. Phys. G {\bf 24}, 1657 (1998).

\bibitem{BertulaniB88}
C.~A. Bertulani and G. Baur, Phys. Rep. {\bf 163},  299  (1988).

\bibitem{MeierHTB98}
H. Meier {\it et~al.}, Eur. Phys. J. C {\bf 2},  741  (1998).

\bibitem{Weinberg97}
S. Weinberg, {\em The Quantum Theory of Fields} (Cambridge University 
Press, Cambridge, 1997), Vol.~1.

\bibitem{AlscherHT97}
A. Alscher, K. Hencken, D. Trautmann, and G. Baur, 
Phys. Rev.~A {\bf 55},  396 (1997).

\bibitem{LandauLQED}
L.~D. Landau and E.~M. Lifschitz, {\em Quantenelektrodynamik}, 
No.~IV in {\em Lehrbuch der theoretischen Physik} 
(Akademie Verlag, Berlin, 1986).

\bibitem{BaurF90}
G. Baur and L.~G. {Ferreira Filho}, Nucl. Phys.~A {\bf 518}, 
786 (1990).

\bibitem{form}
FORM is an algebraical calculation program by J. A. M. Vermaseren. 
The free version 1.0 can be found, e.g., at FTP.NIKHEF.NL.

\bibitem{JacksonED}
J.~D. Jackson, {\em Classical Electrodynamics} 
(John Wiley, New York, 1975).

\bibitem{stagnoliHBT99}
P. Stagnoli, K. Hencken, G. Baur, and D. Trautmann, Differential 
cross sections for QED dielectrons production at relativistic heavy 
ion collisions, abstract submitted to Quark Matter '99, 1999.

\bibitem{FadinK73}
V.~S. Fadin and V.~A. Khoze, Sov. Phys.-JETP {\bf 17},  313  (1973).

\bibitem{JeonKCS98}
S. Jeon, J. Kapusta, A. Chikanian, and J. Sandweiss, 
Phys. Rev.~C {\bf 58}, 1666  (1998).

\end{references}

\end{document}